# $sd^2$ Graphene: Kagome Band in Hexagonal lattice

*Miao Zhou[1], Zheng Liu[1], Wenmei Ming[1], Zhengfei Wang[1], and Feng Liu,[1,2*]*

[1]Department of Materials Science and Engineering, University of Utah, UT 84112, USA.

[2]Collaborative Innovation Center of Quantum Matter, Beijing 100871, China.

*Correspondence to: fliu@eng.utah.edu.*

Graphene, made of $sp^2$ hybridized carbon, is characterized with a Dirac band, representative of its underlying 2D hexagonal lattice. Fundamental understanding of graphene has recently spurred a surge of searching for 2D topological quantum phases in solid-state materials. Here, we propose a new form of 2D material, consisting of $sd^2$ hybridized transition metal atoms in hexagonal lattice, called $sd^2$ 'graphene'. The $sd^2$ graphene is characterized with bond-centered electronic hopping, which transforms the apparent atomic hexagonal lattice into the physics of kagome lattice that may exhibit a wide range of topological quantum phases. Based on first-principles calculations, room temperature quantum anomalous Hall states with an energy gap of ~ 0.1 eV are demonstrated for one such lattice made of W, which can be epitaxially grown on a semiconductor surface of 1/3 monolayer Cl-covered Si(111), with high thermodynamic and kinetic stability.

The structure-property relationships are the most fundamental topics of study in condensed matter and materials physics. One of the most interesting and useful relationships lies between the atomic structure and electronic properties, which often shows clear physical correspondence. For example, in 3D crystalline solids, metals can be viewed as positive ions embedded in a uniform electron gas (jellium model), which form a closely packed fcc/hcp or bcc structure that leads to a metallic band structure. Covalent bonding between group IV elements of C, Si, and Ge leads to formation of an open diamond structure, which in turn leads to a semiconducting band structure with a finite band gap.

Recent emergence of 2D material has revealed some intriguing electronic properties in association with the underlying lattice structure, which shows well-defined one-to-one correspondence. Considering single-orbital nearest-neighbor (NN) hopping, a hexagonal lattice gives rise to linearly dispersive Dirac bands [1], as seen in graphene [2, 3]. The $sp^2$ hybridization in a C hexagonal lattice divides the valence orbitals of C into two sets of bands: one set of $sp^2$ $\sigma$-bands with a large gap and another set of $p_z$ $\pi$-bands within the gap displaying linear dispersion, as illustrated in Fig. 1(a). In presence of spin-orbit coupling (SOC), the Dirac band is also encoded with nontrivial band topology to exhibit quantum spin Hall (QSH) [4] effect. Similarly, a kagome lattice with single-orbital NN hopping gives rise to the so-called kagome band, characterized by a set of Dirac band capped with a flat band [5], as illustrated in Fig. 1(b) (red dots and dashed lines). Both the Dirac and flat bands can be topologically nontrivial in a kagome lattice, leading to a wide range of fascinating topological quantum phases [6, 7].

In this Letter, we propose a new form of 2D material that exhibits a unique kind of structure-property relationship, featured with a hexagonal atomic lattice but kagome electronic bands. Fundamentally, it is a hexagonal lattice consisting of transition metal (TM) atoms with $sd^2$ hybridization, so we name it $sd^2$ 'graphene'. The $sd^2$ hybridization leads to bond-centered σ state hopping on a hexagonal lattice, which can be effectively renormalized into single-orbital NN hopping on a kagome lattice that exhibits typical kagome bands, i.e. a flat band capping two Dirac bands, as illustrated in Fig. 1(b). A specific example is demonstrated, by first-principles calculations, for hexagonal W lattice which can be epitaxially grown on a semiconductor surface of 1/3 monolayer Cl-covered Si(111). We show that beside a typical kagome band, the hexagonal W lattice also displays magnetism and large SOC, supporting room-temperature quantum anomalous Hall (QAH) state with an energy gap as large as 0.1 eV.

Inset of Fig. 2(a) shows the optimized atomic structure of a hexagonal W lattice on the 1/3 monolayer Cl-covered Si(111) surface [W@Cl-Si(111)]. We will first focus on the electronic structures of W@Cl-Si(111) and later discuss its stability and possible experimental growth process. Our electronic structure calculations are based on density functional theory (DFT) and effective Hamiltonian analysis. The details of computational methods are presented in Supplemental Material [8-16]. Despite nonmagnetic nature of bulk W and Si, our calculations show that this surface exhibits ferromagnetism with a magnetic moment of 4 μB per W atom, due to hexagonal crystal field splitting of W $s$ and $d$ orbitals. All spins are found to align in the direction normal to surface, which is crucial for realization of QAH state [17]. The spin-polarized band structures are shown in Figs. 2(a) and (b), with the color coding the relative composition of W and Si (red for W and blue for Si). The energy bands near the Fermi level are mainly contributed by W orbitals, while those of Si are located deeper in energy. Interestingly, this band structure shows a half-metallic behavior: The spin-up component is semiconducting with a gap of 0.15 eV [Fig. 2(a)] and the spin-down component is metallic with two linear bands spreading over the Si bulk gap and a Dirac point formed at the K point that is located exactly at the Fermi level [Fig. 2(b)]. Notably different from the Dirac bands of graphene, the spin-down Dirac bands of W lattice exhibit a strong anisotropy, showing different Fermi velocity along KM *vs* KΓ. There is also a weakly dispersive band on top of the upper Dirac band, and the two bands touch with each other at Γ point. We realize that if this top dispersive band were 'flat', such type of three-band structure would be a typical kagome band [5].

To see more clearly the electronic structure, we plot a 3D band structure of a larger square Brillouin zone containing more high symmetric points, as shown in Fig. 2(c). There are six Dirac cones at the zone boundary (six K points) with dispersion anisotropy clearly seen, while the top weakly dispersive band seems like a cap and touches the upper Dirac band at the Brillouin zone center (Γ). The dispersion anisotropy can be further analyzed by plotting the 2D energy dispersion contour at a constant energy near the Fermi level [Fig. 2(d)]. Interestingly, one sees that at a small doping level of 0.1 eV, the Fermi surface becomes an almost perfect hexagon for both electrons and holes. This nesting feature is associated with the quenched dispersion along the KM direction, which moves the Van Hove singularity close to the Fermi level. For graphene, it has been proposed that the Coulomb interaction in the condition of Van Hove singularity and Fermi surface nesting leads to exotic chiral superconductivity [18]. Similar physics is expected in our case, and should be much easier to be achieved with stronger Coulomb interaction and requiring a lower doping level.

The intrinsic group velocities are further analyzed in association with the anisotropic band dispersion. As shown in Fig. 2(e), the group velocities for both electrons and holes are a strong function of $\vec{K}$. The electron-hole symmetry largely remains with a similar functional shape. The maximum velocity of both is along KΓ direction, and the velocity reduces by ~ 15 % along KM direction. Such group velocity anisotropy will have important consequences for carrier transport [19, 20]. We further found that the degree of anisotropic group velocity depends on substrates, which is different on the W@Br-Si(111) and W@I-Si(111) surfaces (see Fig. S1 in Supplemental Material). This provides a possible way to tune the transport properties of the overlayer W lattice.

To generally understand the physical origin of such peculiar electronic structures, we next provide a detailed orbital analysis associated with an effective Hamiltonian. We first calculated band structures of a 'hypothetical' freestanding hexagonal lattice of W and then use the DFT results as input to construct the maximally localized Wannier functions (WFs) with WANNIER90 code [21]. The resulting WFs are plotted in Fig. 3(a), which show exactly the chemical characteristics of one $s$ and five $d$ orbitals of W. It turns out that the $s$-shape and two $d$-shape WFs with in-plane characteristics ($d_{xy}$ and $d_{x2-y2}$) hybridize to form the bands of interest around the Fermi level. The $sd^2$ hybridization form the bonding σ and antibonding σ* states [see Fig. 3(b)], having the main distribution in the middle of bond. Interestingly, such bond-centered σ states, in contrast to on-site π states in graphene, effectively transform the hexagonal symmetry of atomic W lattice into physics of kagome lattice [Fig. 3(c)]. To capture the most salient features of this transformation, we use single $sd^2$ σ state as the minimum basis to construct an effective Hamiltonian on the transformed bond-centered kagome lattice as,

$$H = -t_1 \sum_{\langle ij \rangle \sigma} c_{i\sigma}^{\dagger} c_{j\sigma} - t_2 \sum_{\langle\langle ij \rangle\rangle \sigma} c_{i\sigma}^{\dagger} c_{j\sigma} - t_3 \sum_{\langle\langle\langle ij \rangle\rangle\rangle \sigma} c_{i\sigma}^{\dagger} c_{j\sigma}$$
$$+ i\lambda \sum_{\langle ij \rangle \alpha\beta} (\vec{E}_{ij} \times \vec{R}_{ij}) \cdot \vec{s}_{\alpha\beta} c_{i\alpha}^{\dagger} c_{j\beta} - M \sum_{i\alpha\beta} c_{i\alpha}^{\dagger} s_{\alpha\beta}^z c_{i\beta}$$
(1),

where $c_{i\sigma}^{\dagger}$ creates an electron with spin σ on lattice site $\vec{r}_i$. ⟨$ij$⟩, ⟨⟨$ij$⟩⟩ and ⟨⟨⟨$ij$⟩⟩⟩ represent the NN, second NN (2NN) and third NN (3NN), respectively. $\vec{R}_{ij}$ is the distance vector between site $i$ and $j$, and $\vec{E}_{ij}$ is the electric field from neighboring ions experienced along $\vec{R}_{ij}$. $\vec{s}$ is spin Pauli matrix. The first three terms describe hopping, the fourth term denotes SOC with magnitude λ and the last term is exchange field with magnitude M. The 2NN and 3NN hopping are included to capture the anisotropy of band dispersion.

Diagonalization of Eq. (1) in reciprocal space for band structure reveals rich underlying physics. As shown in Fig. 3(d), when only the NN hopping is present, the band structure shows a flat band located above two Dirac bands [I in Fig. 3(d)], typical for a kagome lattice [5]. When the 2NN and 3NN hopping are also present, the Dirac bands become anisotropic and the flat band becomes dispersive (II). We found that with $t_1 = 0.46$ eV, $t_2 = 0.018$ eV, $t_3 = -0.016$ eV, and $t_1 = 0.25$ eV, $t_2 = 0.05$ eV, $t_3 = -0.02$ eV, this model can fit very well with the DFT band structures of freestanding W lattice and W@Cl-Si(111) surface, respectively (see Fig. S2). Thus, the main effect of Si substrate is to quench the NN hopping while enhancing the long-range 2NN/3NN hopping.

Upon inclusion of exchange term (M), the spin-up and spin-down bands split. At a considerably large M, the two spin bands are fully separated [III in Fig. 3(d)], leading to a half-metal [22]. Inclusion of SOC will further produce topological quantum phases of the matter. For example, the SOC induced band gap opening at K point of Dirac bands and Γ point between the flat band and upper Dirac will result in QSH state (IV) without magnetism. For system with ferromagnetism and when the exchange field is large enough to overcome the SOC gap, it gives rise to QAH state (V) [22]. Depending on the location of Fermi level, QAH state may exist either in the gap opened at Dirac point (present case) (V) or between the flat band and upper Dirac band. Additionally, the competition between kinetic and Coulomb energy in kagome lattice may lead to rich many-body phases (VI), encompassing superconductor, Mott insulator, magnetically ordered and fractional quantum Hall State, depending on doping level [7]. We point out that all these phase parameters can be possibly controlled, such as via changing the TM elements, pre-adsorbed template atoms/molecules, substrates and applying strain or electrical gating, to realize different quantum phases.

Specifically for the $sd^2$ 'graphene' of W@Cl-Si(111), we found that it supports a QAH surface state with a band gap of 0.1 eV. As shown in Fig.4(a), SOC opens a global gap of ~ 0.1 eV at K point [which is increased slightly to 0.12 eV (Fig. S3) using hybrid functional]. Chern number (C) of all the occupied bands was calculated to be C = -1, indicating existence of one topological transport channel at edges of this surface structure. This is further confirmed by the calculated quantized Hall conductance plateau ($=-e^2/h$) around the Fermi level within the SOC gap [Fig. 4(b)]. We note that to date, QAH effect has yet to be observed in a 2D system. On the other hand, recent experiments [23] have observed QAH effect in magnetic doped 3D topological insulators [24] at very low temperature. Realization of room temperature QAH effect on conventional semiconductor surface is of both scientific and practical interest. Especially, discovering 2D materials that supports room temperature QAH effect is appealing, because it may allow high level experimental control as achieved in conventional Hall measurement of 2D electron gas and graphene.

Next, we discuss possible experimental formation of $sd^2$ graphene. As illustrated in Fig. 5, we suggest a two-step growth process of hexagonal TM lattice on Si substrate: first growing 1/3 monolayer (ML) halogen atoms on Si(111) followed by depositing TM atoms on the pre-adsorbed surface. We have compared different growth structures of Si(111) surface that are partially functionalized by halogen atoms, including F, Cl, Br and I (See details in Sec. IV in Supplemental Material). Particularly at 1/3 ML coverage, while F favors to grow into 'clusters', occupying the NN surface Si sites; Cl, Br and I tend to stay apart from each other forming an 'adatom' superstructure with trigonal symmetry, which is due to the strong steric repulsion between the halogen atoms, as observed in experiments [25-28]. Thus, the 1/3 X-Si(111) (X=Cl, Br, I) surface provides an ideal template for epitaxial growth of hexagonal TM lattices.

Furthermore, W is found to preferably bond with the exposed Si in between Cl and naturally self-assembles into a hexagonal lattice, as guided by the underlying trigonal Cl pattern [Fig. 5(b)]. There is a very strong binding between W and the exposed Si, as evidenced by a typical covalent bond length $d \sim 2.4$ Å and a large adsorption energy ($E_{ad}$) of -6.4 eV. $E_{ad} = E_{W @ Cl-Si(111)} - (E_W + E_{Cl-Si(111)})$, where $E_{W@Cl-Si(111)}$, $E_W$ and $E_{Cl-Si(111)}$ denote the energy of W adsorbed Si(111) surface, W atom, and surface without W, respectively. The adsorption energy of Cl on Si(111) was calculated to be 4.5 eV, suggesting that W is much favored to be adsorbed on the exposed Si site with a dangling bond, gaining 6.4 eV in energy, than to substitute Cl, gaining only $6.4 - 4.5 = 1.9$ eV if Cl desorbs to air or gaining no energy if Cl is displaced to a neighboring Si site. Entropic effects leading to surface substitution and alloying can be effectively suppressed by growth at relatively low temperatures. Moreover, a significant energy barrier of ~ 3.2 eV was found for W atom to diffuse out of the adsorbed Si site to neighboring Cl sites, indicating high kinetic stability of the W-based $sd^2$ graphene in addition to thermodynamic stability.

We stress that it is highly feasible to experimentally realize our proposed two-step epitaxial growth of $sd^2$ graphene, based on the existing related experiments. First, experimental growth of halogen on Si(111) surface is well established in early surface science studies of halogenated Si surface [25-28]. Especially, the 1/3 ML halogen-covered Si(111) surface has been shown to form the trigonal lattice [25-27], exactly as our proposed template structure. Second, previous experiments already indicated that epitaxial growth of metal overlayer on the 1/3 ML halogen-Si(111) template is possible. For instance, indium and Hf hexagonal lattice has been grown on Si(111) $\sqrt{3} \times \sqrt{3}$ -Au surface and Ir(111) substrate, respectively [29, 30]. The only difference is that the surface bands of metal overlayer are strongly overlapping with those of substrates [30, 31], so that this surface does not support topological quantum phase but its growth process can be borrowed for our purpose.

## Acknowledgments

This research was supported by DOE (Grant No: DEFG02-04ER46148). We thank NERSC and the CHPC at University of Utah for providing the computing resources.

## References

1. A. H. Castro Neto, F. Guinea, N. M. R. Peres, K. S. Novoselov, and A. K. Geim, Rev. Mod. Phys. **81**, 109 (2009).

2. K. S. Novoselov, A. K. Geim, S. V. Morozov, D. Jiang, Y. Zhang, S. V. Dubonos, I. V. Grigorieva, and A. A. Firsov, Science **306**, 666 (2004).

3. K. S. Novoselov and A. K. Geim, Nat. Mater. **6**, 183 (2007).

4. C. L. Kane and E. J. Mele, Phys. Rev. Lett. **95**, 226801 (2005).

5. E. Tang, J. W. Mei, and X. G. Wen, Phys. Rev. Lett. 106, 236802 (2011).

6. K. Sun, Z. Gu, H. Katsura, and S. Das Sarma, Phys. Rev. Lett. **106**, 236803 (2011).

7. J. Wen, A. Rüegg, C. C. Wang, and G. A. Fiete, Phys. Rev. B. **82**, 075125 (2010).

8. See Supplemental Material, which includes Refs. [9-16], for computational methods, detailed results on band structures of W@1/3 Br and I-covered Si(111), fitting of DFT results with tight binding kagome model, growth of halogen (F, Cl, Br, I) and organic radicals on Si(111) surface.

9. P. E. Blöchl, Phys. Rev. B **50**, 17953 (1994).

10. G. Kresse and D. Joubert, Phys. Rev. B **59**, 1758 (1999).

11. G. Kresse and J. Hafner, Phys. Rev. B **47**, 558 (1993).

12. G. Kresse and J. Furthmuller, Phys. Rev. B **54**, 11169 (1996).

13. J. P. Perdew, K. Burke, and M. Ernzerhof, Phys. Rev. Lett. **77**, 3865 (1996).

14. J. Heyd, G. E. Scuseria, and M. Ernzerhof, J. Chem. Phys. **118**, 8207 (2003).

15. G. Henkelman, A. Arnaldsson, and H. Jónsson, Comput. Mater. Sci. **36**, 354 (2006).

16. G. Henkelman, B. P. Uberuaga, and H. Jónsson, J. Chem. Phys. **113**, 9901 (2000).

17. H. Zhang, C. Lazo, S. Blügel, S. Heinze, and Y. Mokrousov, Phys. Rev. Lett. **108**, 056802 (2012).

18. R. Nandkishore, L. S. Levitov, and A. V. Chubukov, Nat. Phys. **8**, 158 (2012).

19. C. H. Park, L. Yang, Y. W. Son, M. L. Cohen, and S. G. Louie, Nat. Phys. **4**, 213 (2008).

20. Z. F. Wang and F. Liu, ACS Nano **4**, 2459 (2010).

21. A. A. Mostofi, J. R. Yates, Y. S. Lee, I. Souza, D. Vanderbilt, and N. Marzari, Comput. Phys. Commun. **178**, 685 (2008).

22. Z. F. Wang, Z. Liu, and F. Liu, Phys. Rev. Lett. **110**, 196801 (2013).

23. C. Z. Chang *et al*., Science **340**, 167 (2013).

24. R. Yu, W. Zhang, H. J. Zhang, S. C. Zhang, X. Dai, and Z. Fang, Science **329**, 61 (2010).

25. B. N. Dev, V. Aristov, N. Hertel, T. Thundat, and W. M. Gibson, Surf. Sci. **163**, 457 (1985).

26. S. Rivillon, Y. J. Chabal, L. J. Webb, D. J. Michalak, N. S. Lewis, M. D. Halls, and K. Raghavachari, J. Vac. Sci. Technol., A **23**, 1100 (2005).

27. D. J. Michalak, S. R. Amy, D. Aureau, M. Dai, A. Estève, and Y. J. Chabal, Nat. Mater. **9**, 266 (2010).

28. J. M. Buriak, Chem. Rev. **102**, 1271 (2002).

29. D. V. Gruznev, I. N. Filippov, D. A. Olyanich, D. N. Chubenko, I. A. Kuyanov, A. A. Saranin, A. V. Zotov, and V. G. Lifshits, Phys. Rev. B. **73**, 115335 (2006).

30. L. Li, Y. Wang, S. Xie, X. B. Li, Y. Q. Wang, R. Wu, H. Sun, S. Zhang, and H. J. Gao, Nano Lett. **13**, 4671 (2013).

31. C. H. Hsu, W. H. Lin, V. Ozolins and F. C. Chuang, Phys. Rev. B. **85**, 155401 (2012).

**Figure Captions**

**Fig. 1**. Schematic illustration of structural and electronic properties of (a) $sp^2$ graphene made of C and (b) $sd^2$ graphene made of TM. NN hopping ($t$) of C $p_z$-orbital in a hexagonal lattice (a, left panel) that leads to a Dirac band (a, right panel) is indicated. Also, NN hopping of TM bond-centered $sd^2$ σ state in a hexagonal lattice (b, left panel), which can be renormalized into NN hoping in a kagome lattice (red dots and dashed lines in b, left panel) that leads to a kagome band (b, right panel) is indicated.

**Fig. 2**. Band structures of W@Cl-Si(111) surface for (a) spin-up and (b) spin-down bands. The Fermi level is set at zero. Band compositions are encoded by color: Red for W and blue for Si. Inset of (a) shows the optimized atomic structure of the surface, with purple, green and yellow balls representing W, Cl and Si atoms, respectively. (c) 3D band structures of spin-down component around the Fermi level. Inset shows the surface Brillouin zone. (d) 2D energy dispersion contour of conduction (left panel) and valence band (right panel) cutting at E = ±0.1 eV. (e) Group velocity for electrons and holes measured by the velocity along KΓ direction ($V_k$/$V_{KΓ}$) versus angle ($θ_k$) of the $\vec{K}$ vector. Along KΓ, $θ_K$ = 90°, $V_k$ is maximum; along KM, $θ_k$ = 30°, $V_k$ is minimum, as indicated by two arrows.

**Fig. 3**. (a) The calculated Wannier functions showing the characteristics of *s* and *d* orbitals arranged in order of their energy. (b) Schematics of $sd^2$ hybridization giving rise to the bonding (σ) and antibonding (σ*) state, with the corresponding σ-state Wannier function indicated. (c) Kagome lattice formed by the bond-centered σ state as basis in a hexagonal lattice. The NN, 2NN and 3NN hopping ($t_1$, $t_2$ and $t_3$) are indicated. (d) The model band structure obtained from Eq. 1, illustrating different phases in the parameter space of hopping, magnetization and SOC. Chern number (C) is indicated in V.

**Fig. 4**. (a) Band structures of W@Cl-Si(111) with SOC. The color scheme is the same as in Fig. 3(a). An energy gap of 0.1 eV is opened at K point, with the Chern number of all the occupied bands equal to -1. (b) Anomalous Hall conductance (AHC) around the Fermi level ($E_F$), in units of $e^2/h$.

**Fig. 5**. (a) Schematic view of a two-step epitaxial growth of 2D hexagonal lattices of TM on a semiconductor substrate: First grow 1/3 ML of halogen (X = Cl, Br, I) on Si(111) surface, followed by deposition of TM atoms. (b) and (c) Top and side view of the proposed structure, respectively, with the surface unit cell vector ($a_1$, $a_2$) indicated in (b) and the adsorption length $d$ in (c).

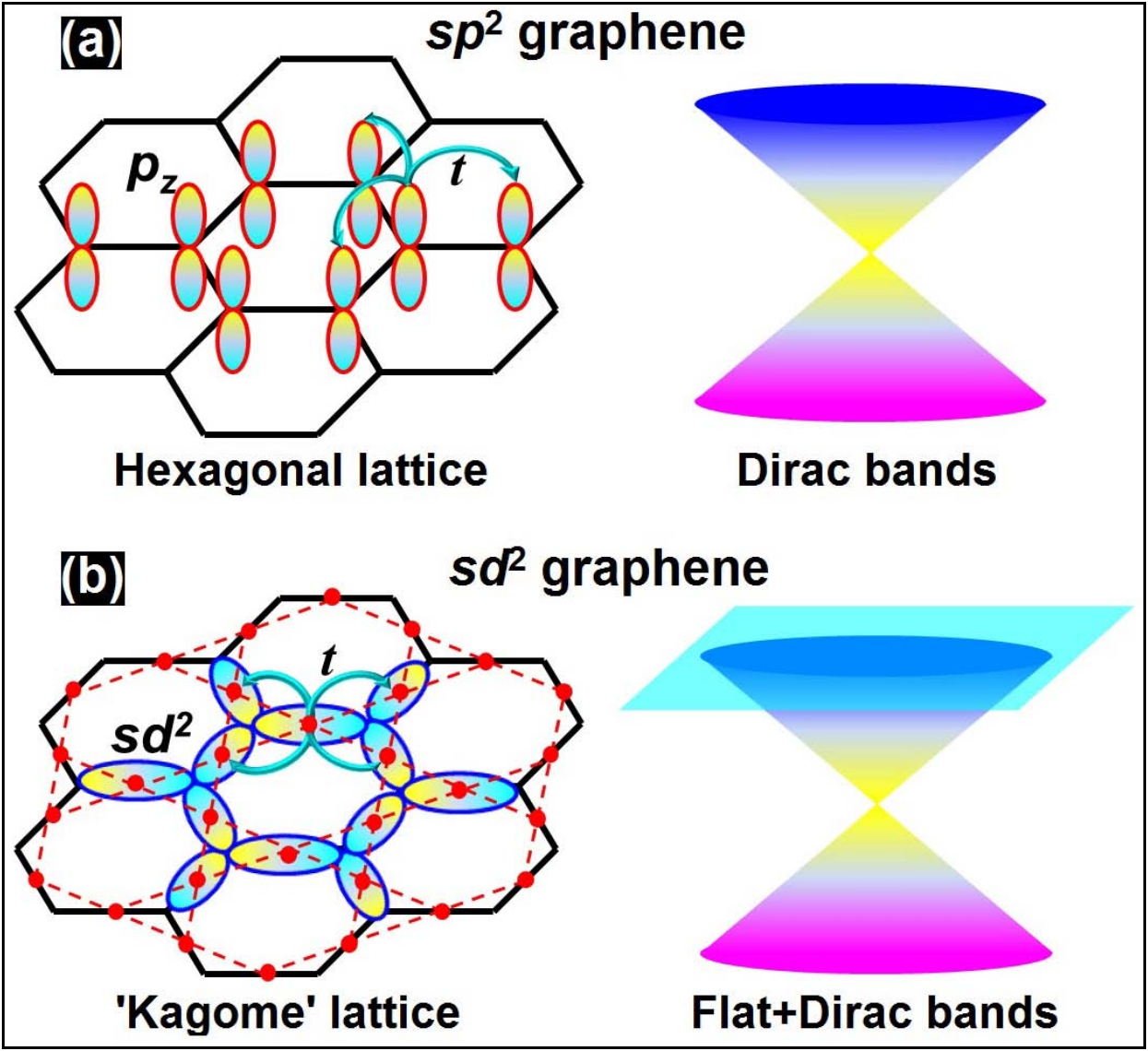

Fig. 1

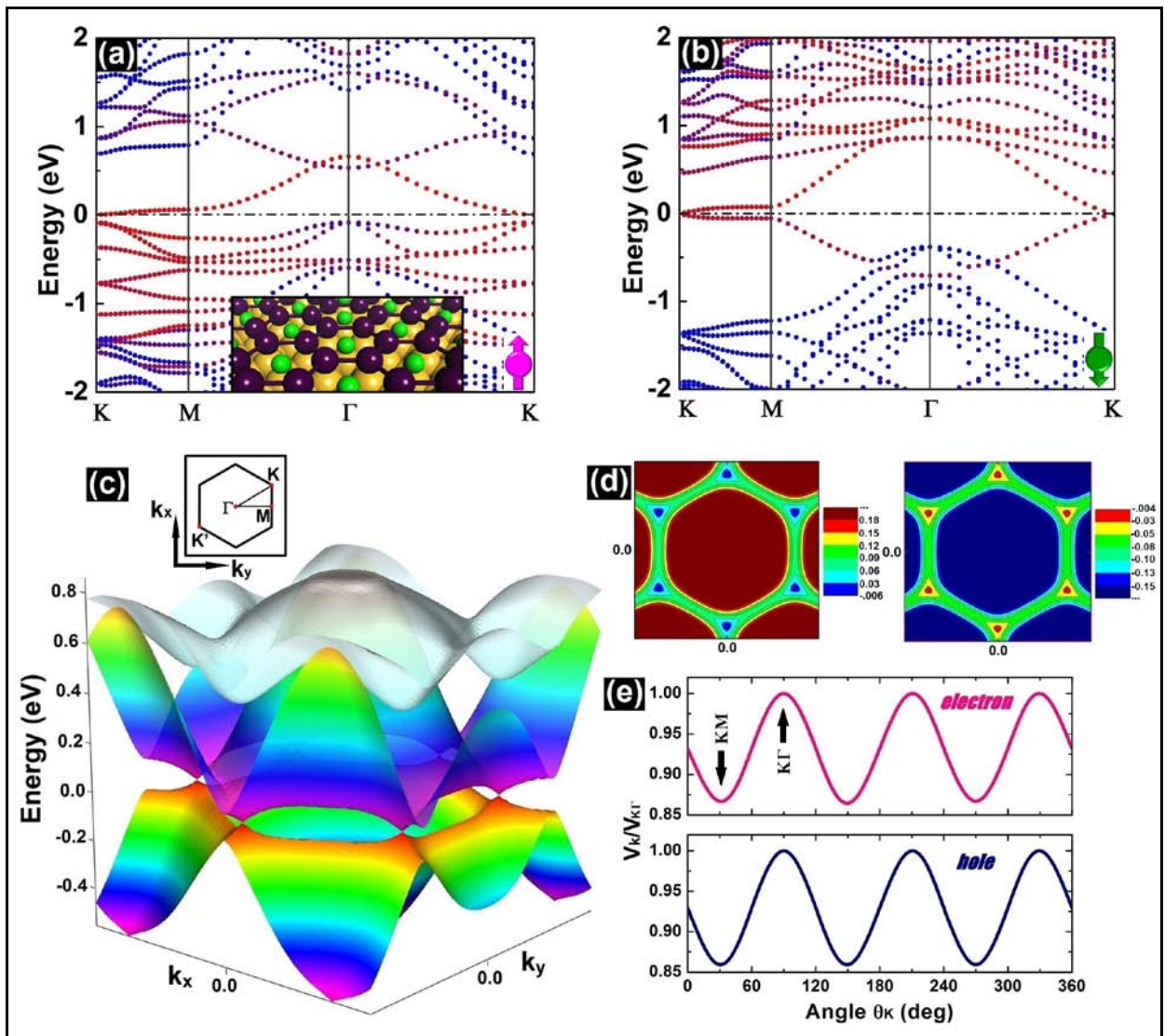

**Fig. 2**

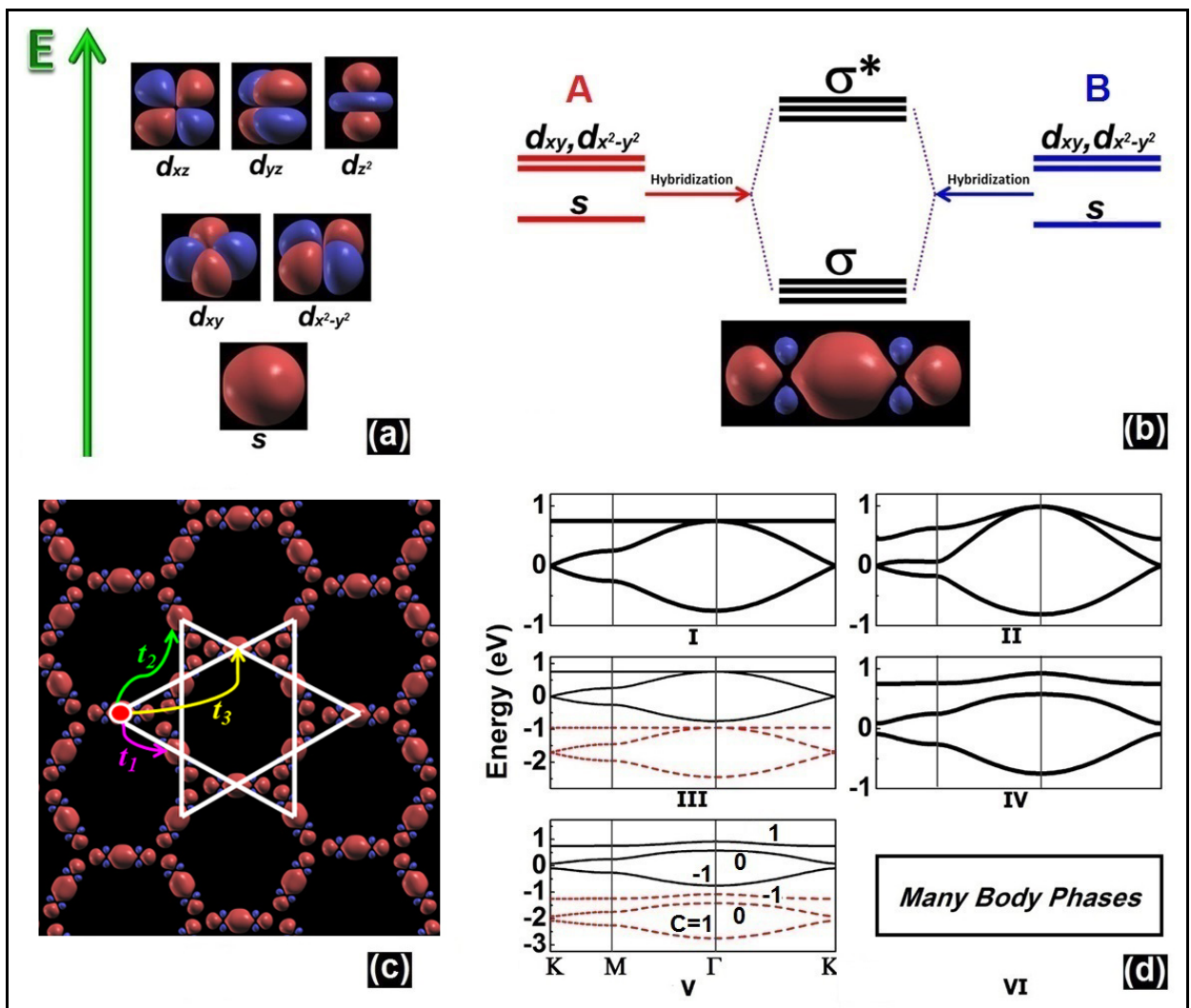

**Fig. 3**

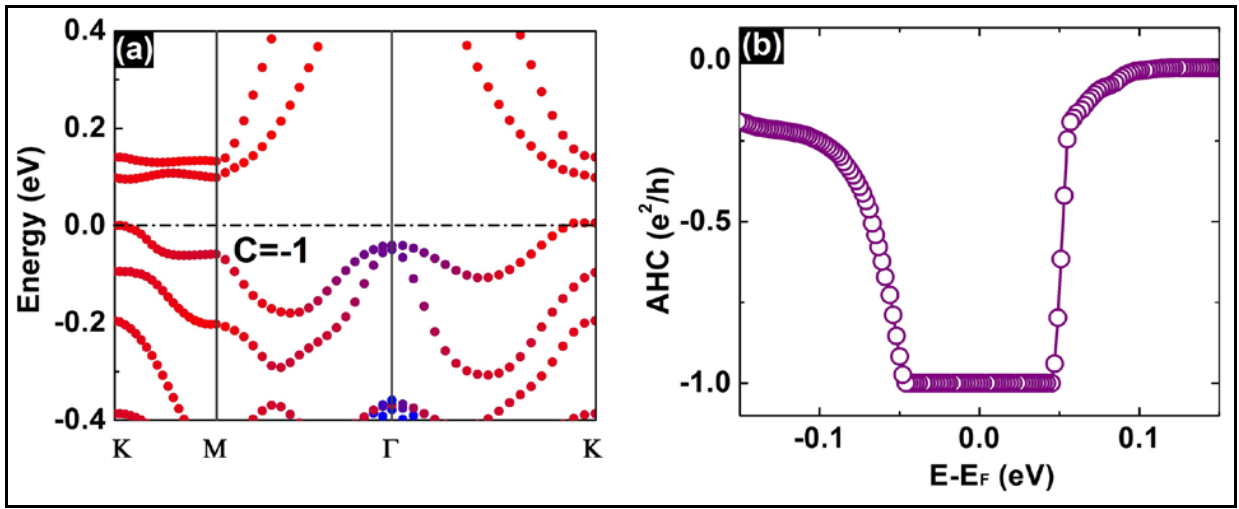

**Fig.4**

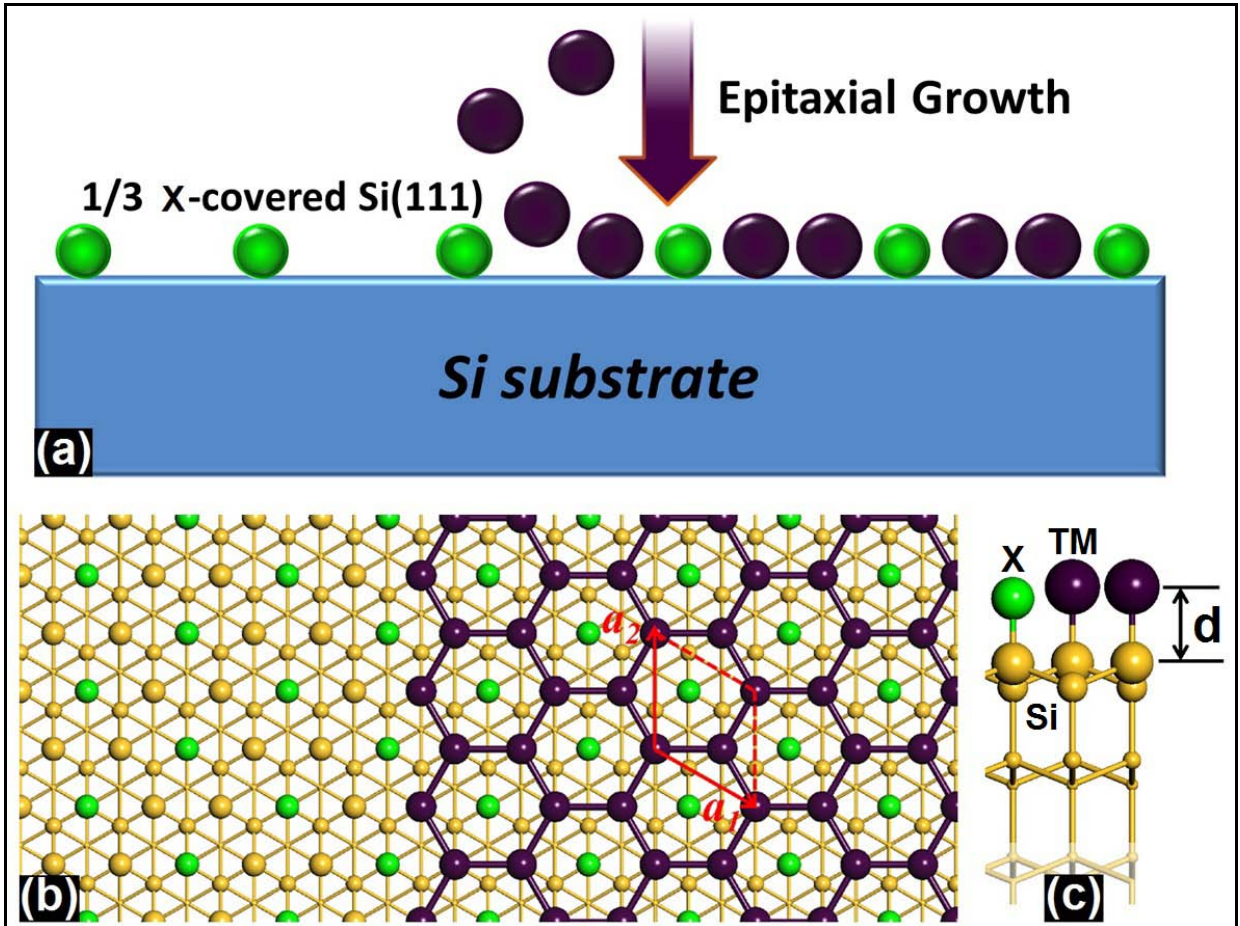

**Fig. 5**

16